\documentclass[aps,prr,reprint,amsmath,amssymb,groupedaddress]{revtex4-2}

\usepackage{graphicx}
\usepackage{dcolumn}
\usepackage{bm}
\usepackage{color}
\usepackage{ulem}
\usepackage{comment}
\newcommand{\ua}{\uparrow}
\newcommand{\da}{\downarrow}
\newcommand{\dg}{\dagger}
\newcommand{\red}[1]{\textcolor{red}{#1}}

\begin{document}


\title{Monte Carlo study of cuprate superconductors in a four-band $d$-$p$ model:\\
Role of orbital degrees of freedom}


\author{Hiroshi Watanabe$^{1}$}
\email{h-watanb@fc.ritsumei.ac.jp}
\author{Tomonori Shirakawa$^{2,3}$}
\author{Kazuhiro Seki$^3$}
\author{Hirofumi Sakakibara$^{4,5,6}$}
\author{Takao Kotani$^{4,5}$}
\author{Hiroaki Ikeda$^7$}
\author{Seiji Yunoki$^{2,3,6,8}$}
\affiliation{
$^1$Research Organization of Science and Technology, Ritsumeikan University, Shiga 525-8577, Japan\\
$^2$Computational Materials Science Research Team, RIKEN Center for Computational Science (R-CCS), Hyogo 650-0047, Japan\\
$^3$Quantum Computational Science Research Team, RIKEN Center for Quantum Computing (RQC), Saitama 351-0198, Japan\\
$^4$Advanced Mechanical and Electronic System Research Center (AMES), Faculty of Engineering, Tottori University, Tottori 680-8552, Japan\\
$^5$Center of Spintronics Research Network (CSRN), Graduate School of Engineering Science, Osaka University, Osaka, 560-8531, Japan\\
$^6$Computational Condensed Matter Physics Laboratory, RIKEN Cluster for Pioneering Research (CPR), Saitama 351-0198, Japan\\
$^7$Department of Physics, Ritsumeikan University, Shiga 525-8577, Japan \\
$^8$Computational Quantum Matter Research Team, RIKEN Center for Emergent Matter Science (CEMS), Saitama 351-0198, Japan
}


\date{\today}

\begin{abstract}
Understanding the various competing phases in cuprate superconductors is a long-standing challenging problem.
Recent studies have shown that orbital degrees of freedom, both Cu $e_g$ orbitals and O $p$ orbitals, are a key ingredient for a unified understanding of cuprate superconductors, including the material dependence.
Here we investigate a four-band $d$-$p$ model derived from the first-principles calculations with the variational Monte Carlo method, which allows us to elucidate competing phases on an equal footing.
The obtained results can consistently explain the doping dependence of superconductivity, antiferromagnetic and stripe phases, phase separation in the underdoped region, and also novel magnetism in the heavily-overdoped region.
The presence of $p$ orbitals is critical to the charge-stripe features, which induce two types of stripe phases with $s'$-wave and $d$-wave bond stripe.
On the other hand, the presence of $d_{z^2}$ orbital is indispensable to material dependence of the superconducting transition temperature ($T_{\mathrm{c}}$), and enhances local magnetic moment as a source of novel magnetism in the heavily-overdoped region as well.
These findings beyond one-band description could provide a major step toward a full explanation of unconventional normal state and high $T_{\mathrm{c}}$ in cuprate supercondutors.
\end{abstract}


\maketitle


\section{INTRODUCTION}\label{intro}
Over 35 years since its discovery~\cite{Bednorz}, cuprate superconductors have continuously challenged our conventional understandings, such as the recent discovery of nematic transitions in the pseudogap region~\cite{Sato}.
It has not yet been achieved to consistently explain the whole phase diagram and the correlation between competing orders and the high transition temperature ($T_{\mathrm{c}}$).
The key to explain the features is considered to be multiorbital effects.
The importance of the orbital degrees of freedom is of great interest in modern condensed matter physics as a source of emergent phenomena such as spin currents~\cite{Murakami,Kato} and, in the field of superconductivity, as a source of novel pairing states and high $T_{\mathrm{c}}$~\cite{Kuroki,Agterberg1}.

In cuprate superconductors, for a long time, the anomalous features have been investigated as the physics of an effective single band crossing a Fermi surface~\cite{Damascelli} rather than multiorbital effects.
The effective one-band models, such as the Hubbard model and the $t$-$J$ model, successfully predicted $d$-wave superconductivity, but were insufficient to describe the material dependence of $T_{\mathrm{c}}$ and the unconventional competing orders.
The $d$-$p$ model or the Emery model~\cite{Emery1}, which consists of Cu $d_{x^2-y^2}$ orbital and O $p_x/p_y$ orbitals, was studied early on as a model involving the multiorbital effect.
These models have been intensively studied in terms of the material dependence of $T_{\mathrm{c}}$, pseudogap phenomena, stripe features, and so on~\cite{Asahata,Takimoto,Yanagisawa,Lorenzana,Shinkai,Kent,Thomale,Arrigoni,Weber1,Weber2,Weber3,Fischer,Weber4,Bulut,Yamakawa,Ogura,White,Huang,Tsuchiizu,Orth,Zegrodnik,Dash,Moreo,Biborski,Cui,Chiciak,Mai}.
Although these models have partially captured the unconventional features, it could not fully explain the anomalous features of cuprates.
For example, the large difference in $T_{\mathrm{c}}$ between $\sim$40 K for a La-based system and $\sim$90 K for a Hg-based system remained unclear.
Recently, one of the authors, Sakakibara and co-authors suggested the importance of the $d_{z^2}$ orbital based on the analysis of a two-orbital model~\cite{Sakakibara}.
The importance of the $d_{z^2}$ orbital has also been supported by the latest angle-resolved photoemission spectroscopy experiment~\cite{Matt}.

These studies indicate that both $p$- and $d$-orbital degrees of freedom should be properly taken into account for the understanding of cuprate superconductors.
In the previous work, we have proposed a four-band $d$-$p$ model composed of the Cu $d_{x^2-y^2}$ and $d_{z^2}$ orbitals and the O $p_x$ and $p_y$ orbitals as a minimal model to obtain the unified description of cuprate superconductors~\cite{Watanabe1}.
On the basis of the variational Monte Carlo (VMC) method, we have shown that this model explains well two key factors about the material dependence of $T_{\mathrm{c}}$:
the contribution of the Cu $d_{z^2}$ orbital to the Fermi surface and the site-energy difference $\Delta_{dp}$ between the Cu $d_{x^2-y^2}$ and O $p$ orbitals~\cite{Ohta}.

In this paper, we investigate the four-band $d$-$p$ model with the VMC method in more detail.
We take the La$_2$CuO$_4$ and HgBa$_2$CuO$_4$ systems as typical examples, and especially, elucidate the competing orders of superconductivity, antiferromagnetic (AF) and stripe phases, phase separation in the underdoped region, and novel magnetism in the heavily-overdoped region.

Our major findings are as follows.
First, charge/spin stripe state is stable over a wide doping range, and its period decreases as the hole doping rate $x$ increases.
At $x=1/8$, we obtain the same stripe phase as observed experimentally, which is extremely robust in the La-based system, but fragile in the Hg-based system.
The AF phase is confined to a narrow doping range near $x\sim 0$.
In the underdoped region $0<x<1/8$, the phase separation (PS) occurs between the AF and the $x=1/8$ stripe phase.
Suppression of the AF correlation by finite intersite $d$-$d$ repulsion causes the transition to a $d$-wave bond stripe observed in non-La-based systems.
Second, concerning the superconductivity, the material dependence of the $T_{\mathrm{c}}$ dome as a function of $x$ is consistently explained.
The dome shape is shown to be strongly correlated with whether the undoped AF insulator is Slater- or Mott-type. 
Finally, we show that novel magnetism observed in the heavily-overdoped region comes from the development of Cu local moments via the Hund’s coupling due to the rapid increase of $d_{z^2}$ component with doping.

The rest of this paper is organized as follows.
In Sec.~\ref{model}, we introduce the model and the numerical method used in this paper.
The four-band $d$-$p$ model on the two-dimensional square lattice is introduced in Sec.~\ref{d-p}.
The VMC method and the Gutzwiller-Jastrow type trial wave function are explained in Sec.~\ref{VMC}.
The tight-binding energy bands for the La$_2$CuO$_4$ and HgBa$_2$CuO$_4$ systems, obtained on the basis of the first-principles calculation, are shown in Sec.~\ref{band structures}.
The numerical results are provided in Sec.~\ref{results}.
The energy competition between AF and stripe phases are studied in detail and the ground state phase diagram including the PS is shown in Sec.~\ref{AFandStripe}.
The symmetry change within the stripe phase is also discussed.
The material and Coulomb interaction dependence of superconductivity is shown in Sec.~\ref{SC} through the behavior of superconducting correlation functions.
The difference between La- and Hg-based systems is discussed from the band structure and electron correlation effects.
The behavior of $d_{z^2}$ hole density and the local magnetic moment is studied in Sec.~\ref{overdoped}.
The novel magnetism in the heavily-overdoped region is also discussed.
Finally, the paper concludes with a summary in Sec.~\ref{Summary}.
The details of the variational wave functions are shown in Appendix~\ref{App}.

\begin{figure}
\centering
\includegraphics[width=0.9\hsize]{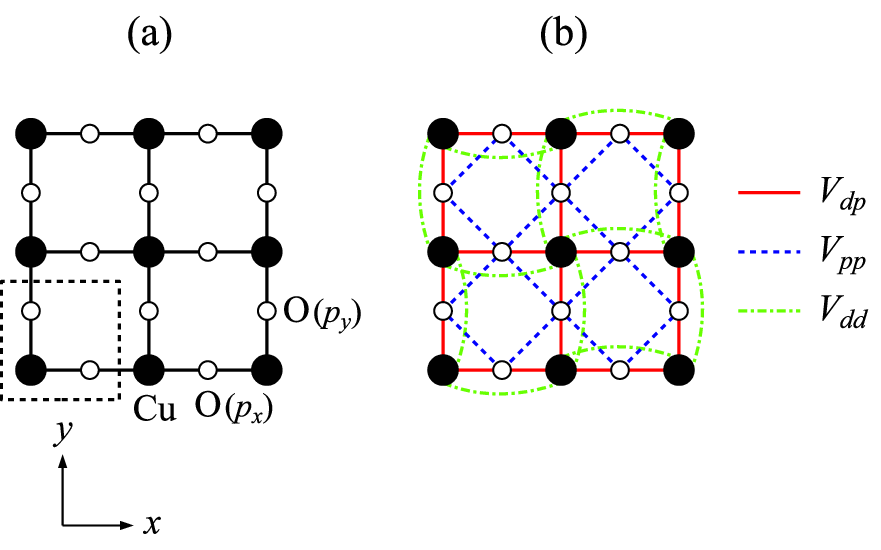}
\caption{\label{lattice}
(a) Lattice structure of the four-band $d$-$p$ model on the two-dimensional square lattice.
There are $d_{x^2-y^2}$ and $d_{z^2}$ orbitals on Cu sites, while there is either $p_x$ or $p_y$ orbital on O sites.
Primitive translation vectors $|\textbf{e}_x|=|\textbf{e}_y|=1$ correspond to the lattice constant between the nearest-neighbor Cu sites.
The dotted square represents a unit cell.
(b) The Coulomb interaction parameters between nearest-neighbor orbitals.
}
\end{figure}

\section{MODEL AND METHOD}\label{model}
\subsection{Four-band $d$-$p$ model}\label{d-p}
As mentioned in Introduction, it is important to incorporate the four orbitals, the Cu $e_g$ orbitals and the O $p_x/p_y$ orbitals, to provide a unified description of the cuprate superconductors. Therefore, we consider a four-band $d$-$p$ model on the two-dimensional square lattice [see Fig.~\ref{lattice}(a)] defined by the following Hamiltonian:
\begin{equation}
H=H_{\text{kin}}+H_{\text{int}}-H_{\text{dc}}.
\end{equation} 
First, the kinetic term $H_{\text{kin}}$ is described by
\begin{align}
H_{\text{kin}}&=\sum_{i,j,\sigma}\sum_{\alpha,\beta}t^{\alpha\beta}_{ij}c^{\dg}_{i\alpha\sigma}c_{j\beta\sigma} \label{kin_orb} \\
                &=\sum_{\textbf{k},\sigma}\sum_m E_m(\textbf{k})a^{\dg}_{\textbf{k}m\sigma}a_{\textbf{k}m\sigma}, \label{kin_band}
\end{align}
where Eq.~(\ref{kin_orb}) is the kinetic term in an orbital representation and Eq.~(\ref{kin_band}) is in a band representation.
$c^{\dg}_{i\alpha\sigma}$ ($c_{i\alpha\sigma}$) is a creation (annihilation) operator of an electron at site $i$ with spin $\sigma\,(=\ua,\da)$ and orbital $\alpha\,(=1,2,3,4)$ 
corresponding to ($d_{x^2-y^2}$, $d_{z^2}$, $p_x$, $p_y$), respectively.
$t^{\alpha\beta}_{ij}$ denotes a hopping integral between orbital $\alpha$ at site $i$ and orbital $\beta$ at site $j$.
$t^{\alpha\alpha}_{ii}=\varepsilon_{\alpha}$ is a site energy for orbital $\alpha$ at site $i$.
These hopping integrals and site energies are determined from the first-principles calculations (see~Sec. \ref{band structures}).
$a^{\dg}_{\textbf{k}m\sigma}$ ($a_{\textbf{k}m\sigma}$) is a creation (annihilation) operator with the wave vector $\textbf{k}$, the energy band index $m\,(=1,2,3,4)$, and spin $\sigma$.
$E_m(\textbf{k})$ is a corresponding energy eigenvalue.

Second, the Coulomb interaction term $H_{\text{int}}$ is composed of eight terms,
\begin{align}
H_{\text{int}}&=U_d\sum_i\left(n^{d_1}_{i\ua}n^{d_1}_{i\da}+n^{d_2}_{i\ua}n^{d_2}_{i\da}\right)+\left(U'_d-\frac{J}{2}\right)\sum_in^{d_1}_in^{d_2}_i  \notag \\
&-2J\sum_i\textbf{S}^{d_1}_i\cdot\textbf{S}^{d_2}_i-J'\sum_i\left(c^{\dg}_{i1\ua}c^{\dg}_{i1\da}c_{i2\ua}c_{i2\da}+c^{\dg}_{i2\ua}c^{\dg}_{i2\da}c_{i1\ua}c_{i1\da}\right) \notag \\
 &+U_p\sum_i\left(n^{p_x}_{i\ua}n^{p_x}_{i\da}+n^{p_y}_{i\ua}n^{p_y}_{i\da}\right)  \notag \\
 &+V_{dp}\sum_{\left<i,j\right>}n^d_in^{p_{x/y}}_j
 +V_{pp}\sum_{\left<i,j\right>}n^{p_x}_in^{p_y}_j+V_{dd}\sum_{\left<i,j\right>}n^d_in^d_j. \label{int}
\end{align}
Here, $n^{\alpha}_i=n^{\alpha}_{i\ua}+n^{\alpha}_{i\da}$ with $n^{\alpha}_{i\sigma}=c^{\dg}_{i\alpha\sigma}c_{i\alpha\sigma}$ is the number operator and $\textbf{S}^{\alpha}_i$ is the spin angular momentum operator at site $i$ with orbital $\alpha$.
$d_1$ and $d_2$ are abbreviations for $d_{x^2-y^2}$ and $d_{z^2}$ orbitals, respectively, and $n^d_i=n^{d_1}_i+n^{d_2}_i$.
$U_d,U_d',J,$ and $J'$ represent on-site intraorbital, interorbital, Hund's coupling, and pair-hopping interactions between $d$ orbitals, respectively.
In this study, we set $J'=J$ and $U_d=U'_d+2J$~\cite{Kanamori}.
$U_p$ is the on-site Coulomb interaction of $p$ orbitals.
$V_{dp},V_{pp},$ and $V_{dd}$ are intersite Coulomb interactions between nearest-neighbor orbitals [see Fig.~\ref{lattice}(b)], where the sum $\sum_{\left<i,j\right>}$ represents nearest-neighbor orbitals located at site $i$ and $j$.
These Coulomb interactions are estimated from the first-principles calculations (see~Sec. \ref{band structures}).

Finally, the double counting correction term $H_{\text{dc}}$ is introduced,
\begin{align}
H_{\text{dc}}&=\biggl[\left\{U_d+2\left(U'_d-\frac{J}{2}\right)+16V_{dd}\right\} \bar{n}^d+8V_{dp}\bar{n}^p\biggr]\sum_in^d_i \notag \\
&+\left\{(U_p+8V_{pp})\bar{n}^p+8V_{dp} \bar{n}^d \right\}\sum_i (n^{p_x}_i+n^{p_y}_i),
\end{align}
where $\bar{n}^d$ and $\bar{n}^p$ are the average electron densities of the $d$ and $p$ orbitals per spin per orbital obtained from the first-principles calculation.
This term is subtracted from the Hamiltonian to correct the energy shift that has already been included in the first-principles calculation.
In the $d$-$p$ model, this double counting correction is important to obtain a reasonable result~\cite{Hansmann}.

\subsection{VMC method}\label{VMC}
In general, it is very difficult to treat the effect of Coulomb interactions correctly.
Here, the VMC method~\cite{McMillan,Ceperley,Yokoyama1} is employed as a powerful computational method.
A Gutzwiller-Jastrow type wave function is considered as a trial state,
\begin{equation}
\left|\Psi\right>=P^{(2)}_{\text{G}}P_{\text{J}_{\text{c}}}P_{\text{J}_{\text{s}}}\left|\Phi\right>.
\label{twf}
\end{equation}
$\left|\Phi\right>$ is a one-body part obtained by diagonalizing the one-body Hamiltonian including many variational parameters and filling the eigen states in ascending order to the corresponding electron density.
The chemical potential is determined in this process.
We can construct long-range-ordered states of charge, spin, and superconductivity.
The explicit forms of them are described in Appendix.

The Gutzwiller factor
\begin{equation}
P^{(2)}_{\text{G}}=\prod_{i,\gamma}\bigl[\text{e}^{-g_{\gamma}}
\left|\gamma\right>\left<\gamma\right|_i\bigr]
\end{equation}
is the one extended for the two-orbital system~\cite{Bunemann,Watanabe2}.
In $P^{(2)}_\text{G}$, possible 16 patterns of charge and spin configuration of the $d_{x^2-y^2}$ and $d_{z^2}$ orbitals at each site $\left| \gamma \right>$, i.e.,
$\left|0\right>=\left|0\;0\right>$, $\left|1\right>=\left|0\ua\right>$, $\cdots$, $\left|15\right>=\left|\ua\da\;\ua\da\right>$,
are differently weighted with $\text{e}^{-g_{\gamma}}$ and $\{g_{\gamma}\}$ are optimized as variational parameters.
The remaining operators
\begin{equation}
P_{\text{J}_{\text{c}}}=\exp\Bigl[-\sum_{i,j}\sum_{\alpha,\beta}v^{\text{c}}_{ij\alpha\beta}n^{\alpha}_in^{\beta}_j\Bigr]
\end{equation}
and
\begin{equation}
P_{\text{J}_{\text{s}}}=\exp\Bigl[-\sum_{i,j}\sum_{\alpha,\beta}v^{\text{s}}_{ij\alpha\beta}s^z_{i\alpha}s^z_{j\beta}\Bigr]
\end{equation}
are charge and spin Jastrow factors, which control long-range charge and spin correlations, respectively.
$\alpha$ and $\beta$ run over four orbitals.
$s^z_{i\alpha}$ is the $z$ component of the spin angular momentum operator at site $i$ with orbital $\alpha$.
The set of $\{v^{\text{c}}_{ij\alpha\beta}\}$ and $\{v^{\text{s}}_{ij\alpha\beta}\}$ are variational parameters.
The variational parameters in $\left|\Psi\right>$ are simultaneously optimized using stochastic reconfiguration method~\cite{Sorella}.
Antiperiodic boundary conditions are imposed on the x- and y-directions of the primitive lattice vectors.

\begin{table*}
\centering
\caption{\label{hopping}
The tight-binding parameters for the La$_2$CuO$_4$ and HgBa$_2$CuO$_4$ systems in eV units.
The definitions of $t_i$ and $\varepsilon_{\alpha}$ are described in Appendix~\ref{NIband}.
}
\renewcommand{\arraystretch}{1.5}
\begin{tabular}{cccccccccc}
    \hline
    \hline
          & $t_1$ & $t_2$ & $t_3$ & $t_4$ & $t_5$ & $t_6$ & $\varepsilon_{d_{x2-y^2}}$ & $\varepsilon_{d_{z^2}}$ & $\varepsilon_{p_{x/y}}$\\
    \hline
       La & 1.42\; & 0.61\; & 0.07\; & 0.51\; & 0.03\; & 0.07 & -0.87 & -0.68 & -3.13  \\
    \hline
       Hg & 1.26\; & 0.65\; & 0.13\; & 0.33\; & 0.00\; & 0.05 & -1.41 & -1.68 & -3.25  \\ 
    \hline
    \hline
\end{tabular}
\renewcommand{\arraystretch}{1}
\end{table*}

\begin{figure*}
\centering
\includegraphics[width=0.8\hsize]{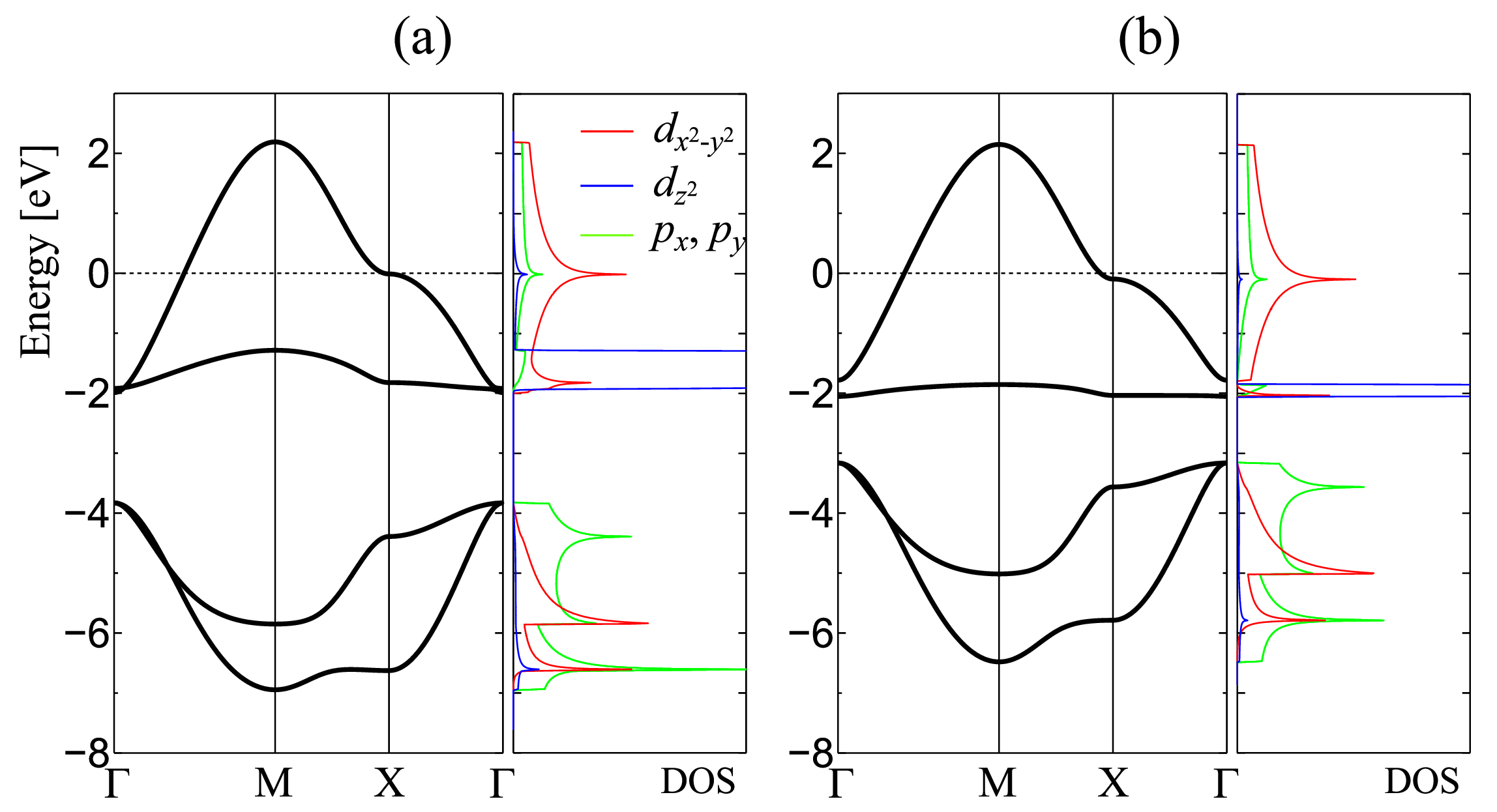}
\caption{\label{band-DOS}
The energy dispersions of the noninteracting tight-binding models for (a) La$_2$CuO$_4$ and (b) HgBa$_2$CuO$_4$ systems.
The projected density of states onto four orbitals are also shown.
Fermi energies correspond to the 15\% hole doping ($x$=0.15).
The high symmetric momenta are indicated as $\Gamma=(0,0)$, $\text{M}=(\pi,\pi)$, and $\text{X}=(\pi,0)$.
}
\end{figure*}

\subsection{Band structures}\label{band structures}
We study the La$_2$CuO$_4$ and HgBa$_2$CuO$_4$ systems as typical examples of single-layer hole-doped cuprates.
The maximally localized Wannier orbitals~\cite{Marzari,Souza} are constructed from the local-density approximation (LDA) with ecalj package~\cite{ecalj}.
The hopping integrals $t_i$ ($i=1-6$) and the site energy of each orbital $\varepsilon_{\alpha}$ are determined to fit the obtained band structure.
They are listed in Table~\ref{hopping} and the explicit form of the tight-binding model is shown in Appendix~\ref{NIband}.
Note that $\varepsilon_{d_{z^2}}$ for the La-based system is corrected to a lower value with reference to the quasiparticle self-consistent $GW$ (QSGW) method~\cite{Faleev,vanShilfgaarde,Kotani3,Jang1}, which gives realistic correction to the LDA.
We have checked its validity in our previous study~\cite{Watanabe1}.

Figure~\ref{band-DOS} shows the noninteracting tight-binding energy bands for La- and Hg-based systems.
We can summarize the difference between them as follows.
(i) The density of states (DOS) of the $d_{z^2}$ component is extended from -1 to -2~eV in the La-based system [Fig.~\ref{band-DOS}(a)], while it is almost localized around -2~eV in the Hg-based system [Fig.~\ref{band-DOS}(b)].
In addition, a small but finite peak structure of the $d_{z^2}$ component exists around the Fermi energy (0~eV) in the La-based system, which greatly affects the stability of superconductivity.
(ii) The site-energy difference
$\Delta_{dp}=\varepsilon_{d_{x^2-y^2}}-\varepsilon_{p}$ is larger in the La-based system (2.26~eV) than in the Hg-based system (1.84~eV).
It affects the strength of the electron correlation through the difference in orbital occupancy.
The small $\Delta_{dp}$ in the Hg-based system leads to a weaker electron correlation because of the more weight of the $p_{x/y}$ orbital.

Starting from these energy band structures, we will investigate the ground state property of the La- and Hg-based systems using the VMC method.
In the following, we set $t_1$ as a unit of energy.
The $d$-orbital on-site Coulomb interaction $U_d/t_1$ is varied as a parameter.
The other Coulomb interaction parameters are set as
$(U'_d,J,U_p,V_{dp},V_{pp})=(0.8, 0.1, 0.5, 0.25, 0.2)\,U_d$ for both La- and Hg-based systems with reference to Ref.~\cite{Hirayama}, unless otherwise noted.
We first set $V_{dd}=0$ and then discuss the effect of finite $V_{dd}$ in Sec.~\ref{symmetry}.

\section{RESULTS}\label{results}
The results obtained are presented in three parts, A. AF and Stripe phases, B. Superconductivity, and C. Heavily-overdoped region.
In VMC calculations, charge ordering and magnetic ordering generally tend to be more stable than superconductivity. This can be due to insufficient incorporation of quantum fluctuations, which is an issue to be addressed in the future.
Instead, the discussion here focuses on the doping and material dependence of each phase.
The effect of randomness is out of the scope of this paper.
We show results for $N_{\text{S}}$=$L\times L$=24$\times$24=576 unit cells (and thus 576$\times$4=2304 orbitals in total), which is large enough to avoid finite-size effects.

\subsection{AF and Stripe phases}\label{AFandStripe}
In this paper, the stripe phase with charge and spin periodicity $\lambda_{\mathrm{c}}$ and $\lambda_{\mathrm{s}}$ is denoted as a C$\lambda_{\mathrm{c}}$S$\lambda_{\mathrm{s}}$ phase according to the convention.
For example, the most familiar stripe phase observed around $x=1/8$ in several La-based systems corresponds to the C4S8 phase.
Our main finding in the charge/spin stripe structure is the ``role-sharing’’ of each orbital; charge modulation occurs mainly on the $p_x$ and $p_y$ orbitals, while spin modulation on the $d_{x^2-y^2}$ orbital.
This is because there is an efficient energy gain due to the orbital degrees of freedom, which is an aspect not present in the one-band model.
Here we show the energetic competition between AF and stripe phases, and discuss the changes in stripe structures along with the doping and material dependence.

\subsubsection{La$_2$CuO$_4$}\label{La2CuO4}
\begin{figure}
\centering
\includegraphics[width=0.8\hsize]{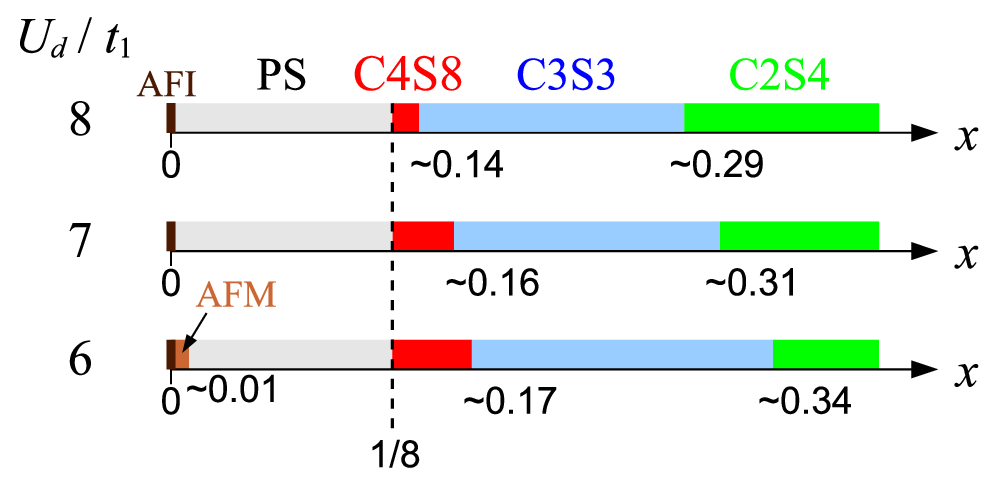}
\caption{\label{pd-La}
Ground state phase diagrams for the La-based system for the hole doping rate $x$ with $U_d/t_1=6$, 7, and 8.
AFI, AFM, and PS represent AF insulator, AF metal, and phase separation, respectively.
C$\lambda_{\mathrm{c}}$S$\lambda_{\mathrm{s}}$ represents the stripe phase with charge and spin periodicity $\lambda_{\mathrm{c}}$ and $\lambda_{\mathrm{s}}$, respectively.
}
\end{figure}

Figure~\ref{pd-La} depicts the ground-state phase diagram obtained in the La-based system as a function of the hole doping rate $x$.
The AF insulator (AFI) at undoped $x=0$ readily becomes unstable with doping.
For large $U_d/t_1$, the AF metallic (AFM) phase does not appear.
For $0<x<1/8$, the PS occurs between the AFI and the C4S8 stripe phase.
Near $x\sim1/8$, the C4S8 stripe phase is quite stable.
The structure of the C4S8 phase is consistent with the experimentally observed $x=1/8$ stripe structure~\cite{Tranquada}.
The stripe phases are widely observed for $x> 1/8$, although the period of the stripes $\lambda_{\mathrm{c}}$ decreases monotonically with further doping.
In the obtained C3S3 and C2S4 stripe phases, the ground-state energies increase drastically. Unlike the C4S8 phase, these stripe phases have no clear reason for their stabilization and also have never been observed experimentally. We believe that these are artificial states of our VMC calculations, and are in fact liquid-like states due to the thermal/quantum fluctuation.
Let us discuss these stripe structure in more detail below.

\begin{figure*}
\centering
\includegraphics[width=0.8\hsize]{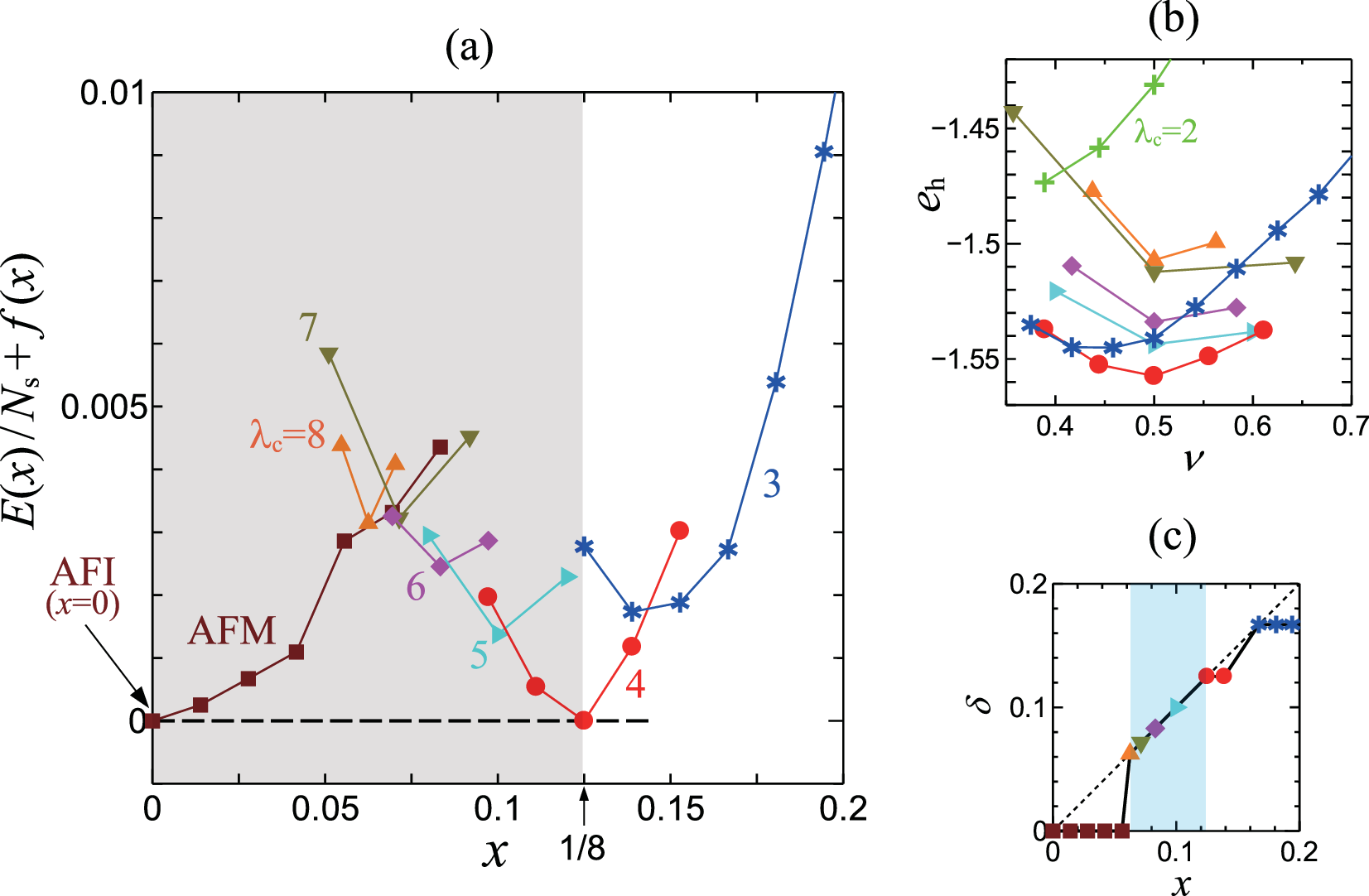}
\caption{\label{ene-La_80}
(a) The $x$ dependence of variational energies per site $E(x)/N_{\mathrm{S}}$ for different phases (AFI, AFM, and stripe phases with charge periodicity $\lambda_{\mathrm{c}}$) for the La-based system with $U_d/t_1=8$.
$f(x)=[-E(0)+8x\{E(0)-E(1/8)\}]/N_{\mathrm{S}}$ is added for visibility.
$t_1=1.42$ eV is a unit of energy.
$N_{\mathrm{S}}=24\times24$ except for $\lambda_{\mathrm{c}}=7$ ($N_{\mathrm{S}}=28\times28$) and $\lambda_{\mathrm{c}}=5$ ($N_{\mathrm{S}}=20\times20$). 
The dashed line represents the Maxwell construction, which indicates the PS for $0<x<1/8$ (gray shaded area).
(b) The excess energy per added hole $e_\mathrm{h}$ for stripe phase as a function of stripe filling $\nu$.
(c) The spin incommensurability $\delta$ as a function of $x$.}
\end{figure*}

Figure~\ref{ene-La_80}(a) shows ground-state energies of AF and stripe phases as a function of $x$.
One can see that AFI and C4S8 phases are fairly stable.
The period of stripe structure $\lambda_{\mathrm{c}}$ monotonically decreases with $x$.
Although the value of $\lambda_{\mathrm{c}}$ is limited to an integer value due to commensurability and finite size effects, it will vary smoothly in the limit of infinite system size.
For $x<1/8$, the PS between AFI and C4S8 phases occurs according to the Maxwell construction, and thus the stripe phase for $\lambda_{\mathrm{c}}>4$ is a metastable state.
The presence of the PS has been reported in previous studies~\cite{Emery2,Arrigoni,Misawa,Ido}. 
In real materials, the effect of lattice distortion is also important for the stability of the stripe phase.
In the La-based systems, such as La$_{2-x}$Sr$_x$CuO$_4$ (LSCO), La$_{2-x}$Ba$_x$CuO$_4$ (LBCO), and La$_{2-x-y}$Nd$_y$Sr$_x$CuO$_4$ (LNSCO), the CuO$_2$ square lattice undergoes a structural transition from a high-temperature tetragonal (HTT) to a low-temperature orthorombic (LTO) phase.
The LTO phase is further deformed to a low-temperature tetragonal (LTT) phase in LBCO and LNSCO. 
Such structural distortion couples to the stripe phases and stabilizes them.
It is possible that some metastable stripe structures in the PS become stable through the structural distortion.


Next, to dissect the stripe structure, we compute the excess energy per added hole~\cite{Lorenzana}, $e_{\mathrm{h}}=\frac{E(x)-E(0)}{x},$ and illustrate it in Fig.~\ref{ene-La_80}(b) as a function of the stripe filling $\nu=x\lambda_{\mathrm{c}}$.
Here, $E(x)$ is the VMC total energy at the hole doping rate $x$.
The stripe filling $\nu$ denotes the hole filling of the folded band structure in the stripe phase.
For instance, $\nu=0.5$ and $\nu=1$ correspond to a half-filled metallic and a fully-filled insulating states, respectively.
In Fig.~\ref{ene-La_80}(b), one can see that the behavior of $e_{\mathrm{h}}$ is drastically changed around $\lambda_{\mathrm{c}}=4$.
For $\lambda_{\mathrm{c}}>4$, $\nu=0.5$ is always stable, especially, the $\lambda_{\mathrm{c}}=4$ (C4S8) phase observed experimentally is the most stable.
This can be the result of maintaining the undoped AFI state as much as possible due to the strong AF correlations, as mentioned in the next paragraph.
Note that in our case, the C8S16 phase with $\nu=1$ proposed in the previous studies does not appear at $x=1/8$.
This phase appears only in peculiar band structures with nearly zero diagonal (next-nearest-neighbor) hopping~\cite{Lorenzana,Ido,Zheng,Jiang}.
It is thus not so realistic.
On the other hand, for $\lambda_{\mathrm{c}}<4$, the stable value of $\nu$ shifts to smaller values as $\lambda_{\mathrm{c}}$ decreases.
Unlike the stripe structure with $\nu=0.5$, there is no clear reason for such a stripe structure to be stabilized.
In fact, stable stripe phases with $\lambda_{\mathrm{c}}<4$ have never been observed experimentally for $x>1/8$.
It is natural to assume that such a stripe state is a fluctuating state like a liquid for $x>1/8$, rather than in a long-range order~\cite{Wen,vonArx}.
More improved calculations incorporating quantum fluctuations can lead to such liquid states~\cite{Huang}.
It is important that such short-range correlations develop for $x>1/8$~\cite{Seibold}, which can have significant effects for the mechanism of high-$T_{\mathrm{c}}$ superconductivity.

Next, let us discuss the spin degree of freedom.
For all the stripe phases considered here, we find the so-called ``spin-charge locking''~\cite{Tranquada,Blackburn}.
Namely, the relation $2\lambda_{\mathrm{c}}=\lambda_{\mathrm{s}}$ ($\lambda_{\mathrm{c}}=\lambda_{\mathrm{s}}$) for even (odd) $\lambda_{\mathrm{c}}$ is always satisfied.
This is because the spin modulation is a driving force of the stripe ordering and the charge modulation only follows it~\cite{Tocchio}.
Indeed, we have checked that the stripe phase without spin modulation, namely, a pure charge stripe phase, is not stabilized within our calculation.
Figure~\ref{ene-La_80}(c) shows the spin incommensurability $\delta$ as a function of $x$.
$\delta$ is defined as the difference from the AF wave vector in $k_x$ direction, $\textbf{k}/2\pi=(\frac{1}{2}\pm\delta,\frac{1}{2})$, where $\textbf{k}$ is a peak position of the spin structure factor of the stripe phase.
In the stripe phases for $\lambda_{\mathrm{c}}\geq 4$, $\delta=x$ is satisfied [blue shaded area for $1/16\leq x\leq 1/8$ in Fig.~\ref{ene-La_80}(c)], which is consistent with ``Yamada relation'' for the La-based system confirmed in neutron scattering experiments~\cite{Yamada}.
This implies that the stripe phases with $\lambda_{\mathrm{c}}\geq 4$ are realized.
We obtain the relation $\delta=x=1/\lambda_{\mathrm{s}}~(1/2\lambda_{\mathrm{s}})$ for even (odd) $\lambda_{\mathrm{c}}$ from these stripe filling $\nu=x\lambda_{\mathrm{c}}=0.5$ and the spin-charge locking.
These results suggest that the origin of the stripe phase in the La-based system is not a band (nesting) effect but a strong correlation effect. Actually, it has been reported that $\lambda_{\mathrm{c}}$ expected from the Fermi-surface nesting has opposite $x$ dependence ($\lambda_{\mathrm{c}}$ increases with increasing $x$)~\cite{Miao}.

\begin{figure}
\centering
\includegraphics[width=0.8\hsize]{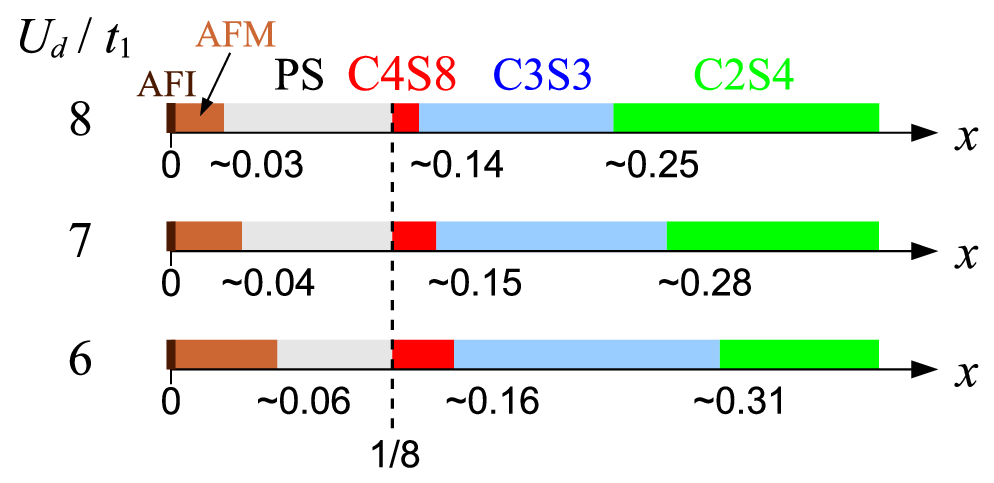}
\caption{\label{pd-Hg}
Ground state phase diagrams for the Hg-based system with $U_d/t_1=6$, 7, and 8.
Notation is the same as in Fig.~\ref{pd-La}.
}
\end{figure}


\subsubsection{HgBa$_2$CuO$_4$}
\begin{figure*}
\centering
\includegraphics[width=0.8\hsize]{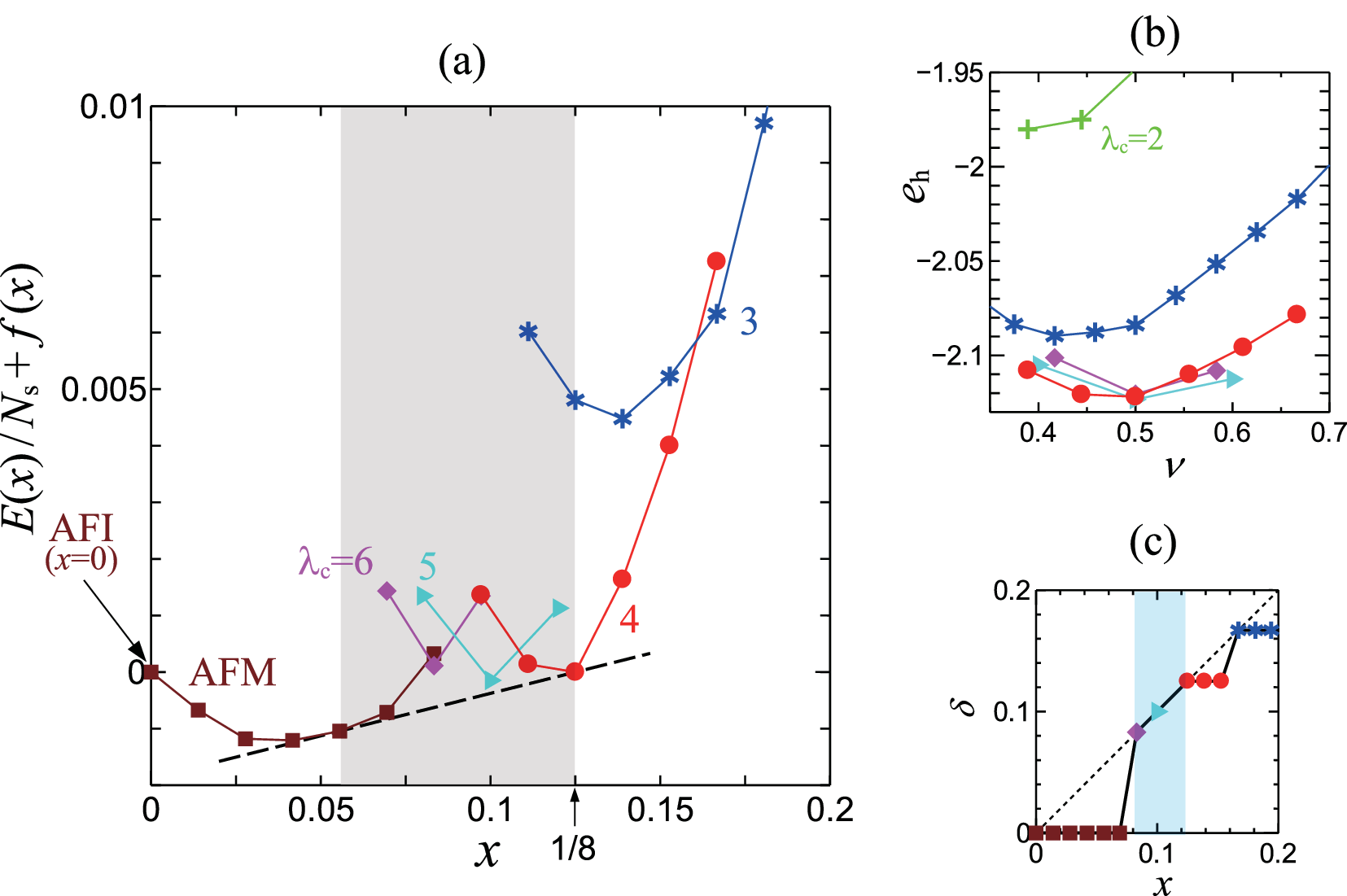}
\caption{\label{ene-Hg_60}
(a) The $x$ dependence of variational energies per site $E(x)/N_{\mathrm{S}}$ for different phases (AFI, AFM, and stripe phases with charge periodicity $\lambda_{\mathrm{c}}$) for the Hg-based system with $U_d/t_1=6$.
$f(x)=[-E(0)+8x\{E(0)-E(1/8)\}]/N_{\mathrm{S}}$ is added for visibility.
$t_1=1.26$ eV is a unit of energy.
$N_{\mathrm{S}}=24\times24$ except for $\lambda_{\mathrm{c}}=5$ ($N_{\mathrm{S}}=20\times20$). 
The dashed line represents the Maxwell construction, which indicates the PS for $0.06<x<1/8$ (gray shaded area).
(b) The excess energy per added hole $e_\mathrm{h}$ for stripe phase as a function of stripe filling $\nu$.
(c) The spin incommensurability $\delta$ as a function of $x$.}
\end{figure*}

Figure~\ref{pd-Hg} shows the ground state phase diagram of the Hg-based system.
Unlike the La-based system, the AFM phase appears in the low-doping region and the PS region is narrower.
As shown in Fig.~\ref{ene-Hg_60}(a), the energy of several metastable stripe phases in the PS region is comparable.
Although the excess energy per added hole, $e_{\mathrm{h}}$, for $\lambda_{\mathrm{c}}\geq 4$ has the minimum at $\nu=0.5$, the energy curve is somewhat shallow compared to the La-based system, as shown in Fig.~\ref{ene-Hg_60}(b).
Therefore, the half-filled stripe phases are not so robust as in the case of La-based system.
This implies that the relation $\delta=x$ for incommensurability [blue shaded area in Fig.~\ref{ene-Hg_60}(c)] is fragile.
In fact, no such relationship has been observed experimentally in the Hg-based systems.
The stripe phases for $x>1/8$ is almost the same as in the La-based system.

\subsubsection{Internal structure of stripe phases and the effect of $V_{dd}$}\label{symmetry}
\begin{figure}[h]
\centering
\includegraphics[width=0.9\hsize]{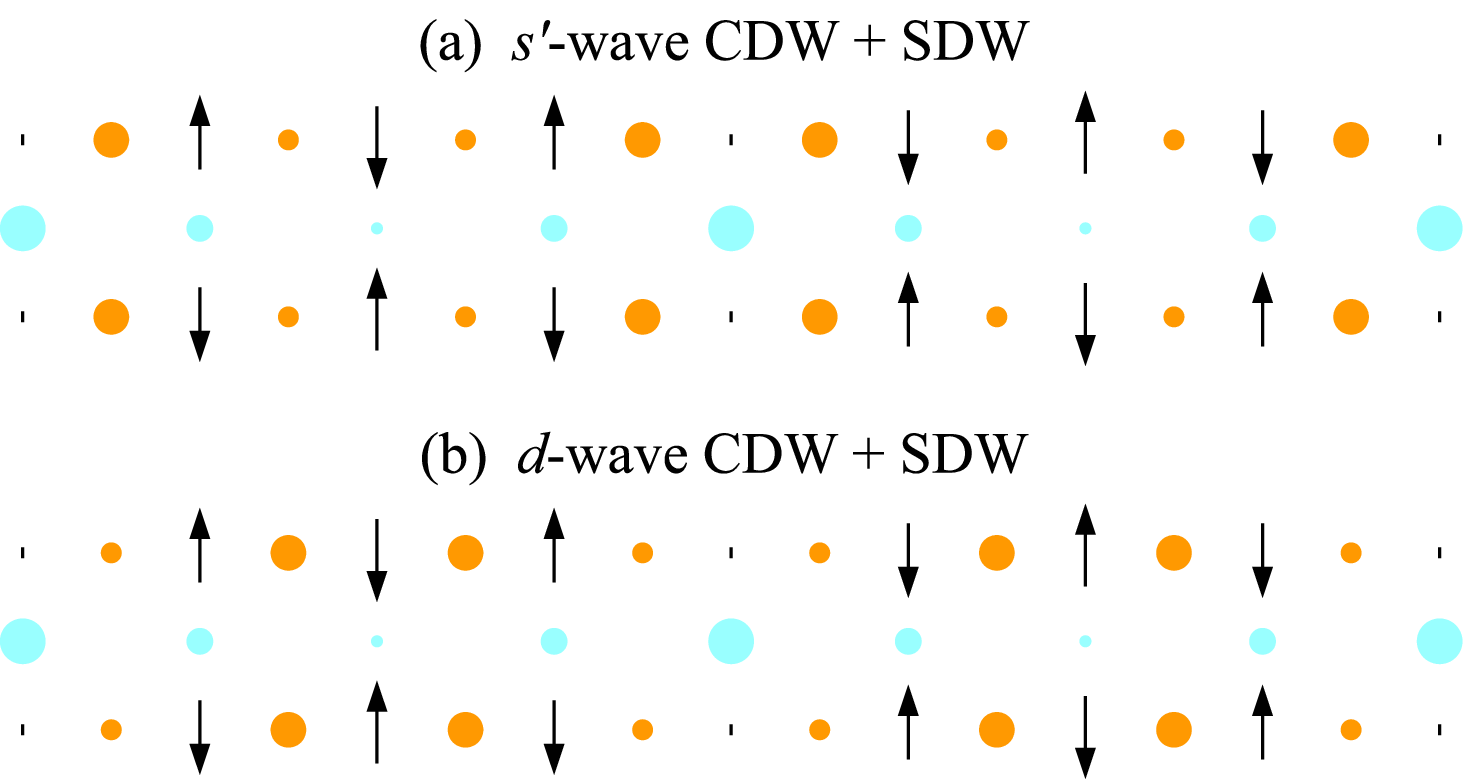}
\caption{\label{stripe-config}
Schematic pictures of the obtained hole and spin density modulations of the C4S8 phase with (a) $s'$-wave CDW+SDW and (b) $d$-wave CDW+SDW.
Orange and blue solid circles represent $p_x$ and $p_y$ holes and each radius is proportional to the hole density.
Arrows represent the $d_{x^2-y^2}$ spins and each length is proportional to the spin density.
}
\end{figure}

Next, let us consider the internal structure of the stripe phases.
Figure~\ref{stripe-config}(a) shows the obtained C4S8 structure.
Reflecting that the cuprates are the charge-transfer insulators, doped hole carriers are mainly introduced into the $p$ orbitals, and the occupation number of the $d$ orbitals remains almost unchanged.
Thus, in Fig.~\ref{stripe-config}, only the spin density is depicted for the $d$ orbitals, while only the hole density is depicted for the $p$ orbitals.
As can be seen in Fig.~\ref{stripe-config}(a), the charge-density wave (CDW) of the doped hole carriers introduced into the $p_x$ and $p_y$ orbitals appears in-phase.
This structure is actually the same as the $s'$-wave bond-order CDW observed in the La-based systems by the resonant soft X-ray scattering experiment~\cite{Achkar}.
For the spin density, the spin-rich Cu sites are surrounded by the hole-poor O sites, and vice versa.
The system efficiently gains the exchange (kinetic) energy around the hole-poor (hole-rich) sites with this configuration.
In this way, spin-active and charge-active areas alternate in a stripe pattern as shown in Fig.~\ref{stripe-config}(a).
The observation of the $s'$-wave CDW is consistent with a slave-boson mean-field approximation~\cite{Lorenzana} and a density-matrix-renormalization-group study~\cite{White} on a three-band $d$-$p$ model.

\begin{figure}[h]
\centering
\includegraphics[width=0.9\hsize]{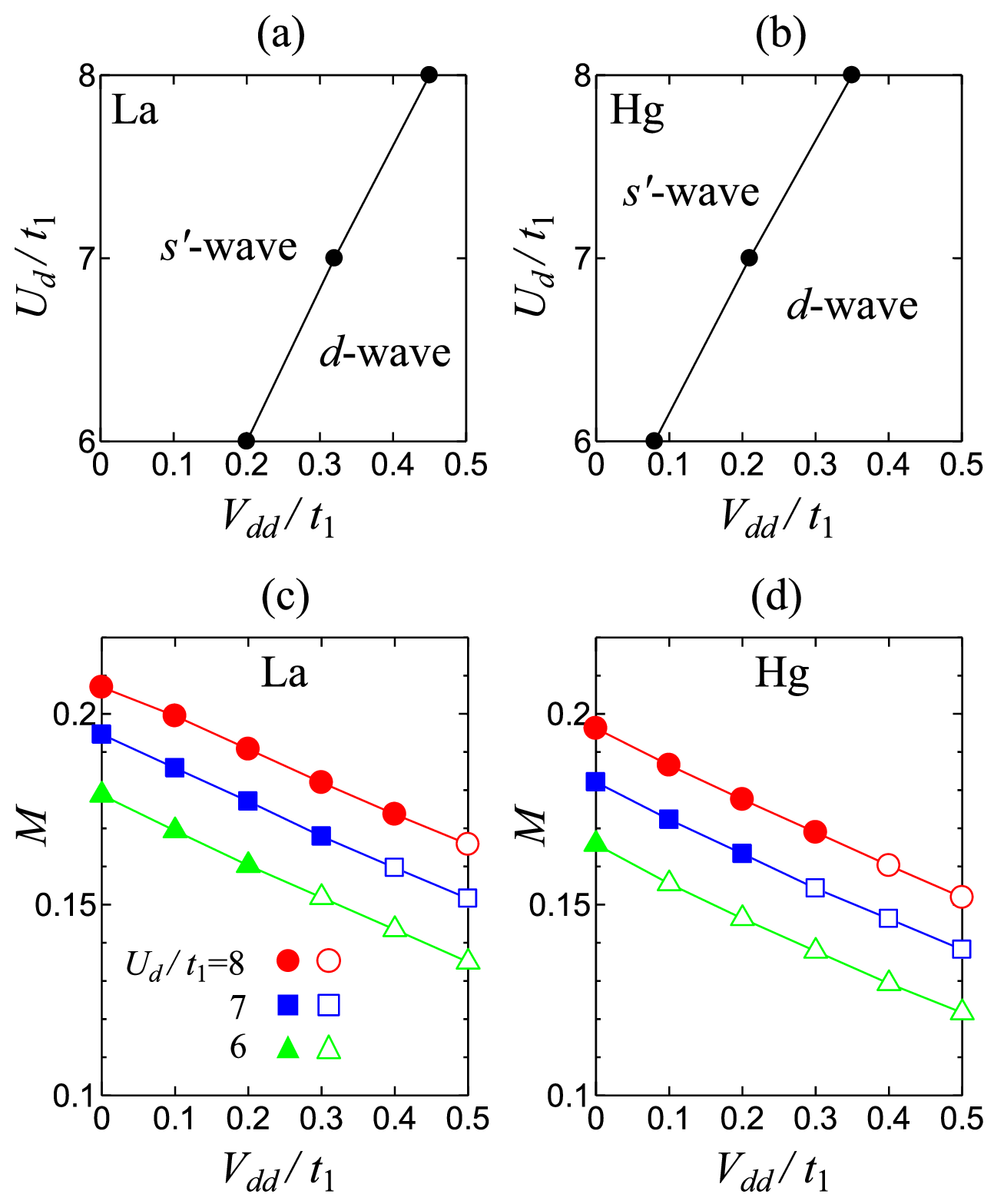}
\caption{\label{s-d}
Internal structure of the C4S8 phase at $x=56/576\sim0.097$ for (a) La-based and (b) Hg-based systems.
The order parameter of the SDW in the C4S8 phase, $M$, for (c) La-based and (d) Hg-based systems.
Solid (open) symbols correspond to the $s'$-($d$-)wave CDW+SDW.
}
\end{figure}

On the other hand, a stripe structure with an anti-phase CDW of the $p$ orbital in Fig.~\ref{stripe-config}(b) (so-called $d$-wave CDW) has been proposed for non-La-based systems, such as Bi$_2$Sr$_2$CaCu$_2$O$_{8+\delta}$ (Bi2212)~\cite{Fujita}, Ca$_{2-x}$Na$_x$CuO$_2$Cl$_2$ (Na-CCOC)~\cite{Fujita}, and YBa$_2$Cu$_3$O$_{6+y}$ (YBCO)~\cite{Comin}.
In the present calculation, the $s'$-wave CDW is stable, but the $d$-wave CDW appears in cases where $V_{pp}$ is a little larger, because $V_{pp}$ has a repulsive effect between neighboring $p_x$ and $p_y$ holes and favors the anti-phase CDW.
We can also confirm the appearance of the $d$-wave CDW by introducing a finite $V_{dd}$, although we have so far assumed $V_{dd}=0$.
As shown in Figs.~\ref{s-d}(a) and~\ref{s-d}(b), the $d$-wave CDW is more likely to appear in the moderately correlated Hg-based systems than in the strongly correlated La-based systems, which is consistent with the experimental results. 

In order to clarify the effect of $V_{dd}$ on the stripe, the magnetization in each state,
\begin{equation}
    M=\sqrt{\frac{S_z(\textbf{q}_{\mathrm{peak}})}{N_{\mathrm{S}}}},
\end{equation}
is shown in Figs.~\ref{s-d}(c) and~\ref{s-d}(d),
where $S_z(\textbf{q}_{\mathrm{peak}})$ represents the peak value of the $z$-component of the $d_{x^2-y^2}$ spin structure factor.
With the introduction of $V_{dd}$, the average magnetization decreases monotonically, and the $d$-wave CDW appears for $M \lesssim 0.16$.
Namely, the introduction of $V_{dd}$ works destructively on the SDW, and eventually the $d$-wave CDW appears.
This is because the $V_{dd}$ effectively weakens $U_d$ by increasing the density of doubly-occupied Cu sites~\cite{Watanabe1}, and thus the system becomes less correlated.  
This situation seems to be a natural connection to the weak correlation approach.
Indeed, it has been argued in several theoretical proposals that the $d$-wave CDW is induced by the nesting effects on the Fermi surface~\cite{Sachdev,Efetov,Yamakawa}.
Since our finite-size calculations have difficulty in capturing the fine structure of the Fermi surfaces, the approach from weak correlations is complementary.

\subsection{Superconductivity}\label{SC}
Next let us consider $d_{x^2-y^2}$-wave spin-singlet superconductivity.
Although the coexistence of stripe and superconducting phases is proposed in several cuprates~\cite{Hamidian,Choubey,Agterberg2}, here we consider the uniform superconducting phase. 
The superconducting correlation function is defined as
\begin{equation}
    P^{dd}(\textbf{r})=\frac{1}{N_{\text{S}}}\sum_i\sum_{\tau,\tau'}f^{(dd)}_{\tau\tau'}\bigl<\Delta^{\dg}_{\tau}(\textbf{R}_i) \Delta_{\tau'}(\textbf{R}_i+\textbf{r})\bigr>,
\end{equation}
where $\Delta^{\dg}_{\tau}(\textbf{R}_i)$ is a creation operator of singlet pairs between nearest-neighbor $d_{x^2-y^2}$ orbitals,
\begin{equation}\label{Delta}
    \Delta^{\dg}_{\tau}(\textbf{R}_i)=\frac{1}{\sqrt{2}}(c^{\dg}_{i1\uparrow}c^{\dg}_{i+\tau1\downarrow}+c^{\dg}_{i+\tau1\uparrow}c^{\dg}_{i1\downarrow}),
\end{equation}
and $\tau$ represents four nearest-neighbor Cu sites ($\tau=\pm \mathbf{e}_x,\pm\mathbf{e}_y$).
$f^{(dd)}_{\tau\tau'}$ is a form factor of a superconducting gap function with $d_{x^2-y^2}$ symmetry, namely,
$f^{(dd)}_{\tau\tau'}=1$ for $\tau \parallel \tau'$ and $-1$ for $\tau \perp \tau'$.
If $P^{dd}(\textbf{r})$ is saturated to a finite value for $r=|\mathbf{r}|\rightarrow\infty$, superconducting long-range order exists.
In the following, we take the saturated value of $P^{dd}(r\rightarrow\infty)$ as a strength of superconductivity $P^{dd}$.

\begin{figure*}
\centering
\includegraphics[width=0.8\hsize]{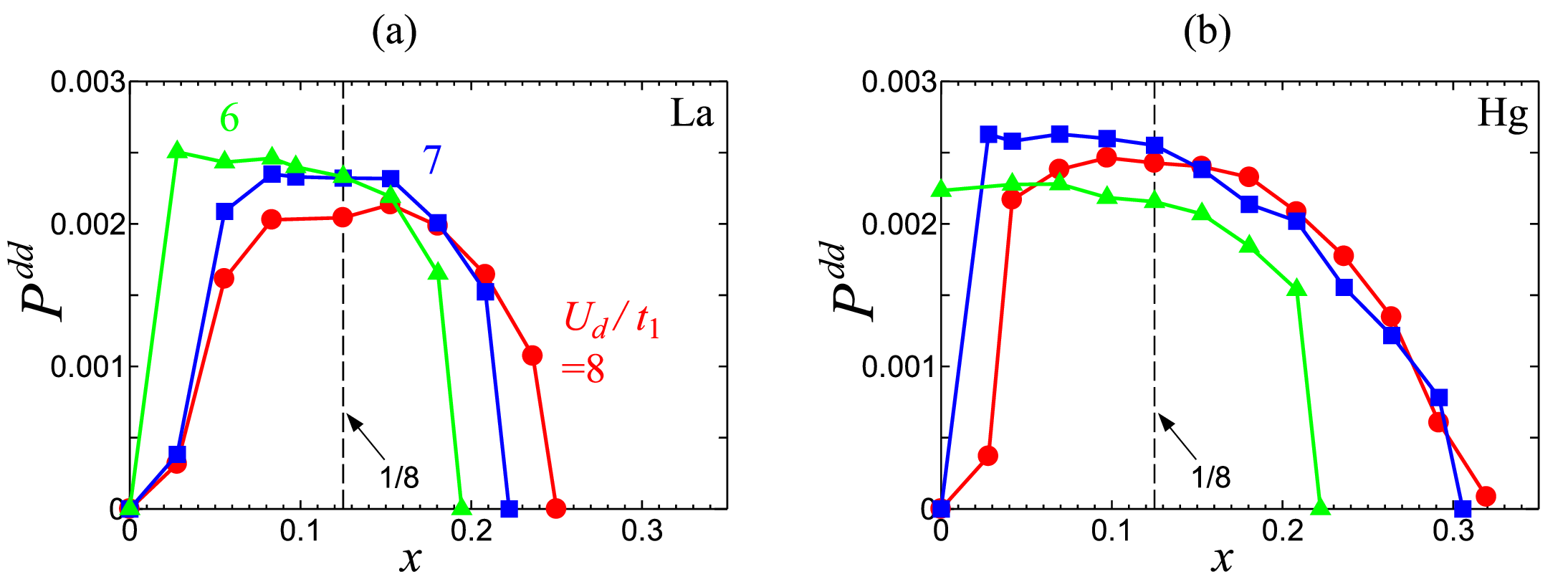}
\caption{\label{Pdd}
The $x$ dependence of superconducting correlation function $P^{dd}$ for (a) La-based and (b) Hg-based systems.
$N_{\mathrm{S}}=24\times24$.
}
\end{figure*}

The $x$ dependence of $P^{dd}$ for the La-based system is shown in Fig.~\ref{Pdd}(a).
At $x=0$, the system is an insulator due to correlation effects, and thus superconductivity is completely suppressed, $P^{dd}=0$.
As $x$ increases, the mobility of the Cooper pairs increases due to the introduction of mobile carriers by doping.
On the other hand, the strength of the $d$-$d$ pairing itself is reduced by doping due to the reduction of electron correlation.
The former effect is predominant in a low-doping, strongly-correlated region and the latter effect is predominant in a high-doping, weakly-correlated region.
In other words, the peak position of the dome corresponds to a boundary between strongly- and weakly-correlated regions.
Therefore, the peak position is moved to a smaller value of $x$ with decreasing $U_d/t_1$ because the weakly-correlated region is extended.

The $x$ dependence of $P^{dd}$ for the Hg-based system is shown in Fig.~\ref{Pdd}(b).
While the dome-shaped behavior is observed for $U_d/t_1=7$ and 8, $P^{dd}$ monotonically decreases with $x$ for $U_d/t_1=6$.
This is because for $U_d/t_1=6$, the system is not insulating but metallic at $x=0$ when the paramagnetic state is assumed.
When the AF order is taken into account, the crossover from the Mott insulator to the Slater insulator occurs at $U_d/t_1\sim6.3$ in the present model~\cite{Watanabe1}.
Namely, the view of a ``doped Mott insulator'' is no longer valid for $U_d/t_1=6$.
The shape of $P^{dd}$ can be a measure of the strength of an electron correlation.
From this point of view, the La-based system is more strongly correlated than the Hg-based system, because the system is still a Mott insulator at $x=0$ for $U_d/t_1=6$, as shown in Fig.~\ref{Pdd}(a).
It results from the larger value of $\Delta_{dp}$ in the La-based system, which leads to a larger $d$-orbital occupancy of holes and a stronger electron correlation~\cite{Watanabe1}.

We can observe that $P^{dd}$ of the Hg-based system is larger than that of the La-based system.
It is consistent with the higher $T_{\mathrm{c}}$ in the Hg-based system ($\sim$90~K) compared with the La-based system ($\sim$40~K).
The reason why $P^{dd}$ of the Hg-based system is larger can be understood as follows:
(i) The low $\varepsilon_{d_{z^2}}$ in the Hg-based system is favorable for superconductivity through the less $d_{z^2}$-orbital contribution to the Fermi surface;
the $d_{z^2}$-orbital contribution to the Fermi surface is destructive for superconductivity due to the localized character of the $d_{z^2}$ electrons~\cite{Watanabe1}.
This effect is significant for a high-doping ($x\gtrsim 0.15$) region.
(ii) The small $\Delta_{dp}(>0)$ in the Hg-based system is favorable for superconductivity in a low-doping region ($x\lesssim 0.15$) through the weaker electron correlation due to the hybridization of $p$ orbitals;
the electron correlation is destructive for superconductivity in a low-doping region because the mobility of the Cooper pairs is suppressed.
$P^{dd}$ itself is a physical quantity in the ground state but is closely related to $T_{\mathrm{c}}$.
We expect that the larger $P^{dd}$, the higher $T_{\mathrm{c}}$.
We consider that both the band structure and electron correlation effects are necessary to explain the material dependence of $T_{\mathrm{c}}$.

We also note that the dip structure around $x=1/8$ becomes slightly more visible as $U_d/t_1$ increases, in both La-based and Hg-based systems. 
In fact, the dip structure around $x=1/8$ is widely observed in cuprate superconductors.
This structure is more pronounced the more stable the stripe states are, as in LBCO and LNSCO with the LTT distortion mentioned in Sec.~\ref{La2CuO4}.
The dip structure we found is tightly related to this fact.
This implies that although our variational wave functions for the superconductivity do not explicitly include the charge modulation, stripe fluctuations are partially included through the Jastrow factors.

Finally, we mention the finite-size effect.
The dome-shaped behavior and the material dependence of $P^{dd}$ are obtained also for $N_{\mathrm{S}}=16\times16$.
We consider that $N_{\mathrm{S}}=24\times24$ used here is large enough to discuss the superconductivity in the present model.

\subsection{Heavily-overdoped region}\label{overdoped}
\begin{figure*}
\centering
\includegraphics[width=0.8\hsize]{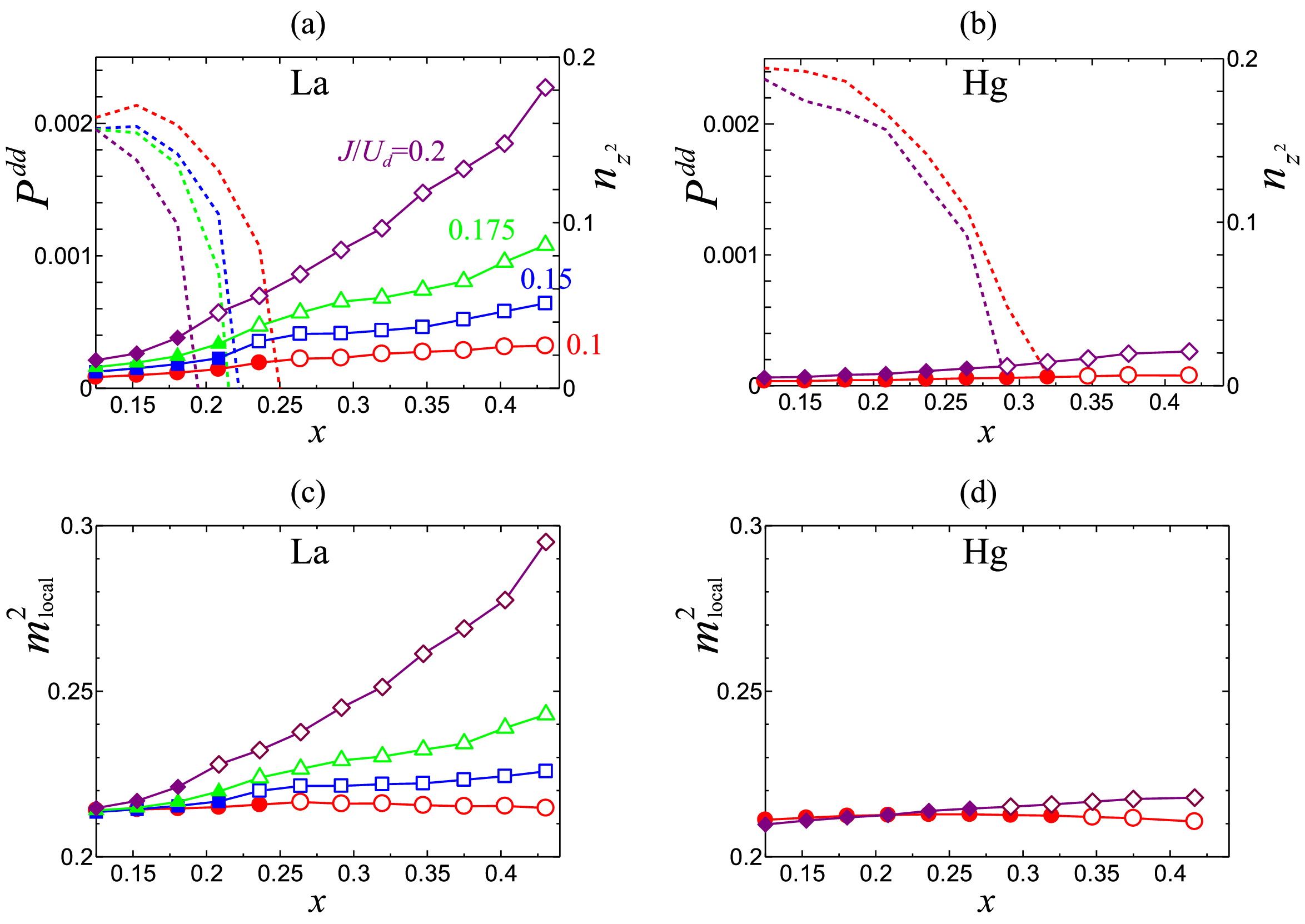}
\caption{\label{Hund}
The $x$ dependence of superconducting correlation function $P^{dd}$  (dotted lines) and the average hole density of the $d_{z^2}$ orbital $n_{z^2}$ (solid lines with symbols) for (a) La-based and (b) Hg-based systems.
The $x$ dependence of local moment $m_{\mathrm{local}}$ for (c) La-based and (d) Hg-based systems.
Solid (open) symbols correspond to the superconducting (paramagnetic) phase.
The colors indicate several $J/U_d$ values.
}
\end{figure*}

Here, we discuss the hole density of the $d_{z^2}$ orbital $n_{z^2}$ and novel magnetism in the heavily-overdoped region.
First of all, in Figs.~\ref{Hund}(a) and~\ref{Hund}(b), we illustrate $n_{z^2}$ as a function of $x$ for several $J/U_d$ at $U_d/t_1=8$ in the La-based and Hg-based systems, along with the superconducting correlation function $P_{dd}$.
In the La-based system, $n_{z^2}$ increases almost linearly with $x$, and is strongly enhanced by the Hund's coupling $J/U_d$.
Such enhancement of $n_{z^2}$ in the heavily-overdoped region has been also reported by the high-resolution Compton scattering experiment in LSCO~\cite{Sakurai}.
Note that $n_{z^2}$ is suppressed in a finite $P_{dd}$ region. 
The negative correlation between $P_{dd}$ and $n_{z^2}$ suggests that the $d_{z^2}$ orbital works destructively for superconductivity, which is consistent with the previous studies~\cite{Sakakibara,Watanabe1}.
Such superconducting-elastic effect could explain the intriguing pressure effect where $c$-axis compression reduces superconducting $T_{\mathrm{c}}$~\cite{Hardy}, because the $c$-axis compression leads to the increase of $n_{z^2}$ through lowering the apical oxygen height.
On the other hand, in the Hg-based system, 
such an increase of $n_{z^2}$ and suppression of $P_{dd}$ are not so pronounced even at relatively large $J/U_d=0.2$.
This is because the $\varepsilon_{d_{z^2}}$ is much lower in the Hg-based system and the $d_{z^2}$ orbital is almost inactive. 

Second, in Figs.~\ref{Hund}(c) and~\ref{Hund}(d), we show the local moment $m_{\mathrm{local}}$, 
\begin{equation}
    m_{\mathrm{local}}=\sqrt{\left<S^z_iS^z_i\right>}
\end{equation}
where $S^z_i=\sum_{\alpha}S^z_{i\alpha}$ is the $z$ component of the total spin angular momentum operator at site $i$ and $\alpha$ runs over four orbitals in the unit cell.  
As shown in Figs.~\ref{Hund}(c) and~\ref{Hund}(d), $m^2_{\mathrm{local}}$ exhibits quite similar behavior to $n_{z^2}$.
In the La-based system, it suggests that the spin of doped holes in the $d_{z^2}$ orbital is ferromagnetically aligned with that of the $d_{x^2-y^2}$ orbital due to the Hund's coupling, which leads to the increase of $m_{\mathrm{local}}$.
Namely, new magnetic ``seeds'' appear in the heavily-overdoped region originated from the $d_{z^2}$ orbital degree of freedom, which is absent in the one-band Hubbard and even in the three-band $d$-$p$ models.
On the other hand, in the Hg-based system, such a multiorbital effect is not observed because the $d_{z^2}$ orbital is almost inactive.
The relative difference in the position of the $d_{z^2}$ orbital causes a clear difference in the doping dependence of the La and Hg systems.

Finally, let us comment a novel magnetism in the heavily-overdoped region.
It has been discussed in various ways.
Magnetic susceptibility measurements suggest the emergence of local paramagnetic moments in LSCO~\cite{Wakimoto} or ferromagnetic spin fluctuations in (Bi,Pb)$_2$Sr$_2$CuO$_{6+\delta}$ (Bi-2201)~\cite{Kurashima,Komiyama}.
A zero-field $\mu$SR measurement in LSCO also suggests the development of ferromagnetic spin fluctuations~\cite{Sonier}.
Theoretical studies show the possibility of ferromagnetism using a quantum scaling theory~\cite{Kopp}, first-principles calculations for supercells of LBCO~\cite{Barbiellini}, determinant quantum Monte Carlo~\cite{Jia}, dynamical cluster approximation~\cite{Maier}, and fluctuation exchange approximation~\cite{Teranishi}.
We can consider two mechanisms for the ferromagnetism in the heavily-overdoped region.
One is the band effect.
The peak position of the spin susceptibility at $\textbf{q}=(\pi,\pi)$ in the low-doping region moves toward $\textbf{q}=(0,0)$ with hole doping.
Indeed, the Lifshitz transition where the topology of the Fermi surface changes from hole-like to electron-like occurs around $x\sim 0.3$ in the one-band Hubbard model with $t'/t=-0.3$, leading to the enhancement of $\chi_0(0,0)$. 
The other is the multiorbital effect, i.e., the double-exchange mechanism originated from the Hund's coupling between the $d_{x^2-y^2}$ and $d_{z^2}$ orbitals.
Our result supports this mechanism.
We expect that these two mechanisms will not contradict but rather cooperatively work toward ferromagnetism.

\section{DISCUSSION AND SUMMARY}\label{Summary}

In this paper, we present a unified explanation of the doping and material dependence of the ground state in cuprate superconductors using the VMC analysis of the four-band $d$-$p$ model derived from the first-principles calculations.
First, we found the following results from a comparison of total energies of the AFI, AFM, and stripe phases. 
In the Hg-based system, a slight hole-doping causes the appearance of AFM phase, but in the La-based systems, the AFM state is less likely to appear due to strong correlations.
In the La-based system, the experimentally observed $s'$-wave bond stripe phase is quite stable, and all stable stripe structures are in the spin-charge locking state.
Overall features of spin incommensurability observed in the neutron scattering experiments are explained.
In the Hg-based system, introducing a small but finite $V_{dd}$ or slightly increasing $V_{pp}$ can lead to a transition to the $d$-wave bond stripe structure and even a spin-charge unlocked state.
PS appears in the low-doping region of both La- and Hg-based systems.
All of these behaviors are consistent with experimental observations, and our results generally capture their characteristic features well.

Next, we computed the superconducting correlation functions $P^{dd}$, and clarified that the dome-shaped $P^{dd}$, indicative of the $T_{\mathrm{c}}$ dome as a function of the hole doping rate, is originated from the dual nature of the Coulomb repulsion, that is, the enhancement of the pairing interaction and the depairing effect by the decrease of mobile electron pairs. 
$\varepsilon_{d_{z^2}}$ and $\Delta_{dp}$ are material-dependent key parameters for high $T_{\mathrm{c}}$.
Our result suggests that deep $\varepsilon_{d_{z^2}}$ and small $\Delta_{dp}$ raise the $T_{\mathrm{c}}$. 
This is the reason why $T_{\mathrm{c}}$ of the Hg-based system is higher than that of the La-based system.

Finally, we investigated novel magnetism in heavily-overdoped region. 
For the hole doping rate $x>1/8$, both the $d_{z^2}$ hole density $n_{z^2}$ and the local moment at Cu sites increases almost linearly with $x$ in the La-based system. 
This means that the $e_g$ orbital spins are aligned parallel via the Hund's coupling.
These local spins can be ferromagnetically aligned by the double-exchange mechanism, which is consistent with previous proposals.
On the other hand, such a multiorbital effect is not observed in the Hg-based system because the $d_{z^2}$ orbital is located much lower and thus almost inactive.

These findings, which are beyond one-band description, could provide a major step toward a full explanation of unconventional normal state and high-$T_{\mathrm{c}}$ in cuprate supercondutors.

\begin{acknowledgments}
The authors thank K. Kuroki, T. Tohyama, and T. Adachi for useful discussions.
The computation has been done using the facilities of the Supercomputer Center, Institute for Solid State Physics, University of Tokyo and the supercomputer system HOKUSAI in RIKEN.
This work was supported by a Grant-in-Aid for Scientific Research on Innovative Areas ``Quantum Liquid Crystals'' (KAKENHI Grant No. JP19H05825) from JSPS of Japan,
and also supported by JSPS KAKENHI (Grant Nos. JP22K03520, JP22K03512, JP21H04446, JP20K03847, JP19K23433, JP19H01842, and JP18H01183).

\end{acknowledgments}

\appendix*
\section{Construction of the trial wave function}\label{App}

\subsection{Noninteracting energy band}\label{NIband}
First, we show the construction of the noninteracting tight-binding energy band discussed in Sec.~\ref{band structures}.
It is obtained by diagonalizing the following one-body Hamiltonian:
\begin{widetext}
\begin{align}
H_{\text{kin}}&=\sum_{\textbf{k},\sigma}
        \left( c_{\textbf{k}1\sigma}^{\dg}, c_{\textbf{k}2\sigma}^{\dg},  
   c_{\textbf{k}3\sigma}^{\dg}, c_{\textbf{k}4\sigma}^{\dg}\right) 
 \begin{pmatrix}
   t_{11} & t^*_{21} & t^*_{31} & t^*_{41} \\ 
   t_{21} & t_{22} & t^*_{32} & t^*_{42} \\
   t_{31} & t_{32} & t_{33} & t^*_{43} \\
   t_{41} & t_{42} & t_{43} & t_{44}
 \end{pmatrix}
 \begin{pmatrix}
  c_{\textbf{k}1\sigma} \\ c_{\textbf{k}2\sigma} \\ 
  c_{\textbf{k}3\sigma} \\ c_{\textbf{k}4\sigma}    
 \end{pmatrix} \label{kin_band_app1}\\
      &=\sum_{\textbf{k},\sigma}\sum_m E_m(\textbf{k})a^{\dg}_{\textbf{k}m\sigma}a_{\textbf{k}m\sigma} \label{kin_band_app2}
\end{align}
with the hopping matrix elements given as
\begin{align}
t_{11}&=\varepsilon_{d_{x^2-y^2}},\\
t_{21}&=0,\\
t_{22}&=\varepsilon_{d_{z^2}}-2t_5(\cos k_x+\cos k_y),\\
t_{31}&=2\text{i}t_1\sin\frac{1}{2}k_x,\\
t_{32}&=-2\text{i}t_4\sin\frac{1}{2}k_x,\\
t_{33}&=\varepsilon_{p_x}+2t_3\cos k_x+2t_6[\cos(k_x+k_y)+\cos(k_x-k_y)],\\
t_{41}&=-2\text{i}t_1\sin\frac{1}{2}k_y,\\
t_{42}&=-2\text{i}t_4\sin\frac{1}{2}k_y,\\
t_{43}&=2t_2\left[\cos\left(\frac{1}{2}k_x+\frac{1}{2}k_y\right)-\cos\left(\frac{1}{2}k_x-\frac{1}{2}k_y\right)\right],\\
t_{44}&=\varepsilon_{p_y}+2t_3\cos k_y+2t_6[\cos(k_x+k_y)+\cos(k_x-k_y)],
\end{align}
\end{widetext}
where $c^{\dg}_{\textbf{k}\alpha\sigma}$ ($c_{\textbf{k}\alpha\sigma}$) is a creation (annihilation) operator of an electron with wave vector $\textbf{k}$, spin $\sigma\,(=\ua,\da)$, and orbital $\alpha\, (=1,2,3,4)$
corresponding to ($d_{x^2-y^2}$, $d_{z^2}$, $p_x$, $p_y$), respectively.
The hopping integrals $t_i$ ($i=1-6$) and the site energy of each orbital $\varepsilon_{\alpha}$ are determined to fit the band structures obtained from the LDA or QSGW calculation.

\subsection{Trial wave function}
\subsubsection{Superconductivity}
To construct the trial wave function for superconductivity, we employ the Bogoliubov de-Gennes (BdG) type Hamiltonian in real space~\cite{Himeda}, i.e., 
\begin{equation}
H_{\text{BdG}}=\sum_{i,j}\sum_{\alpha,\beta}\left(c^{\dg}_{i\alpha\ua}, c_{i\alpha\da}\right)
 \begin{pmatrix}
  T^{\alpha\beta}_{ij\ua} & \Delta^{\alpha\beta}_{ij} \\ 
  \Delta^{\alpha\beta*}_{ji} & -T^{\alpha\beta}_{ji\da}
 \end{pmatrix}
 \begin{pmatrix}
  c_{j\beta\ua} \\
  c^{\dg}_{j\beta\da}
 \end{pmatrix}.
\end{equation}
Here, $T^{\alpha\beta}_{ij\sigma}$ is a normal part and corresponds to  the 4$\times$4 matrix in Eq. (\ref{kin_band_app1}) with renormalized hopping integrals $\tilde{t}_i$ and also includes the chemical potential term.
The chemical potential $\mu$ is set to the corresponding Fermi energy.
$\Delta^{\alpha\beta}_{ij}$ is an anomalous part that represents the superconducting pairing in real space.
Therefore, the variational parameters to be optimized in $\left|\Phi\right>$ are $\tilde{t}_i$ ($i=2-6$) and $\{\Delta^{\alpha\beta}_{ij}\}$ with $\tilde{t}_1=t_1$ being fixed as a unit of energy.
In this study, the pairing between nearest-neighbor orbitals, $d$-$d$, $d$-$p_x$, $d$-$p_y$, $p_x$-$p_x$, and $p_y$-$p_y$, are considered. $d$ denotes the $d_{x^2-y^2}$ orbital.
If we set $\Delta^{\alpha\beta}_{ij}=0$, the paramagnetic phase is obtained.


\subsubsection{Uniform spin AF and stripe phases}
As mentioned in Sec.~\ref{VMC}, various long-range orderings of charge and spin can be described by introducing $\{\rho^{\alpha}_i\}$ and $\{m^{\alpha}_i\}$.
A uniform spin AF phase along the $z$- and $x$-direction can be introduced as
\begin{equation}
  m^{\alpha}_{(z)}\sum_i e^{i\textbf{Q}\cdot\textbf{r}_i}\left(c^{\dg}_{i\alpha\ua}c_{i\alpha\ua}-c^{\dg}_{i\alpha\da}c_{i\alpha\da}\right)
\end{equation}
and
\begin{equation}
  m^{\alpha}_{(x)}\sum_i e^{i\textbf{Q}\cdot\textbf{r}_i}\left(c^{\dg}_{i\alpha\ua}c_{i\alpha\da}+c^{\dg}_{i\alpha\da}c_{i\alpha\ua}\right)
\end{equation}
for each orbital $\alpha\,(=1,2,3,4)$, where $\textbf{Q}=(\pi,\pi)$, and $m^{\alpha}_{(z)}$ and $m^{\alpha}_{(x)}$ are treated as variational parameters.
They are energetically degenerate within a one-body description.
However, they will give different variational energies with the Gutzwiller and Jastrow factors in Eq.~(\ref{twf}) that break the SU(2) symmetry.
We have confirmed that $m^{\alpha}_{(x)}$ always gives a lower variational energy and thus a better trial state.
This is because the Gutzwiller and Jastrow factors generate spin fluctuations in the direction orthogonal to that of $m^{\alpha}_{(x)}$.

For a stripe phase with charge and spin periodicities $\lambda^{\alpha}_{\text{c}}=2\pi/q^{\alpha}_{\text{c}}$ and $\lambda^{\alpha}_{\text{s}}=2\pi/q^{\alpha}_{\text{s}}$, respectively, the following potentials with spatial modulation in the $x$ direction should be introduced at site $i$ for each orbital $\alpha\,(=1,2,3,4)$:
\begin{equation}
\rho^{\alpha}_i=\rho^{\alpha}\cos[q^{\alpha}_{\text{c}}(x_i-x^{\alpha}_{\text{c}})]
\end{equation}
and
\begin{equation}
m^{\alpha}_i=(-1)^{x_i+y_i}m^{\alpha}\sin[q^{\alpha}_{\text{s}}(x_i-x^{\alpha}_{\text{s}})],
\end{equation}
where $\rho^{\alpha}$ and $m^{\alpha}$ are the amplitude of charge and spin orderings, respectively.
$x^{\alpha}_{\text{c}}$ and $x^{\alpha}_{\text{s}}$ control the relative phases of charge and spin orderings, respectively.
$\lambda^{\alpha}_{\text{c}}$, $\lambda^{\alpha}_{\text{s
}}$, $\rho^{\alpha}$, $m^{\alpha}$, $x^{\alpha}_{\text{c}}$ and $x^{\alpha}_{\text{s}}$ are all variational parameters to be optimized.

\end{document}